\documentclass[aps,twocolumn,superscriptaddress,nofootinbib,a4paper,longbibliography]{revtex4-2}
\pdfoutput=1

\usepackage{times}
\usepackage{epsfig}
\usepackage{amsfonts}
\usepackage{amsmath}
\usepackage{amsthm}
\usepackage{amssymb}					
\usepackage{amsthm}
\usepackage{dsfont}
\usepackage{bm}
\usepackage{mathtools}
\usepackage{color}
\usepackage{multirow}
\usepackage[normalem]{ulem}
\usepackage{latexsym}
\usepackage{mathrsfs}
\usepackage{natbib}
\usepackage{verbatim}
\usepackage[T1]{fontenc}
\usepackage{graphicx}
\usepackage{xcolor}
\usepackage{physics}
\usepackage{textcomp}
\usepackage{tikz-cd}

\usepackage{floatrow}
\usepackage{subfig}

\usepackage{caption}
\usepackage[colorlinks=true,linkcolor=blue,citecolor=magenta,urlcolor=blue]{hyperref}
\newcommand{\bracket}[3]{\langle#1|#2|#3\rangle}

\begin{document}


\title{Almost qudits in the prepare-and-measure scenario}

\author{Jef Pauwels}
\affiliation{Laboratoire d'Information Quantique, CP 225, Universit\'e libre de Bruxelles (ULB), Av. F. D. Roosevelt 50, 1050 Bruxelles, Belgium}
\author{Stefano Pironio}
\affiliation{Laboratoire d'Information Quantique, CP 225, Universit\'e libre de Bruxelles (ULB), Av. F. D. Roosevelt 50, 1050 Bruxelles, Belgium}
\author{Erik Woodhead}
\affiliation{Laboratoire d'Information Quantique, CP 225, Universit\'e libre de Bruxelles (ULB), Av. F. D. Roosevelt 50, 1050 Bruxelles, Belgium}
\author{Armin Tavakoli}
\affiliation{Institute for Quantum Optics and Quantum Information - IQOQI Vienna, Austrian Academy of Sciences, Boltzmanngasse 3, 1090 Vienna, Austria}
\affiliation{Atominstitut,  Technische  Universit{\"a}t  Wien, Stadionallee 2, 1020  Vienna,  Austria}

\begin{abstract}
Quantum communication is often investigated in scenarios where only the dimension of Hilbert space is known. However, assigning a precise dimension is often an approximation of what is actually a higher-dimensional process. Here, we introduce and investigate quantum information encoded in carriers that nearly, but not entirely, correspond to standard qudits. We  demonstrate the relevance of this concept for semi-device-independent quantum information by showing how small higher-dimensional components can significantly compromise the conclusions of established protocols. Then we provide a general method, based on semidefinite relaxations, for bounding the set of almost qudit correlations, and apply it to remedy the demonstrated issues. This method also offers a novel systematic approach to the well-known task of device-independent tests of classical and quantum dimensions with unentangled devices. Finally, we also consider viewing almost qubit systems as a physical resource available to the experimenter and determine the optimal quantum protocol for the well-known Random Access Code.
\end{abstract}


\maketitle


\textit{Introduction.---} The Hilbert space dimension of a system is a key property in quantum theory. Most experiments assume knowledge of it because it reflects the number of relevant independent degrees of freedom. Indeed, even the fundamental unit of quantum information, namely the qubit, is expressed in terms of the (minimal meaningful) quantum dimension. It is natural that much research has been devoted to the quantum dimension: the device-independently certification of it \cite{Brunner2008, Gallego2010, Hendrych2012, Ahrens2012, Pauwels2021}, investigating the cost of classically simulating qubits \cite{Galvao2003, Toner2003, Renner2022}, using $d$-dimensional quantum systems (qudits) to outperform $d$-dimensional classical systems (dits) in useful tasks \cite{Nayak1999, Navascues2015, Tavakoli2015} and performing quantum information protocols in experiments where nothing but the dimension is assumed to be known \cite{Pawlowski2011, Woodhead2015,  Tavakoli2018, Tavakoli2020}. A large number of experiments have followed  (see e.g.~\cite{Bennet2014, Mironowicz2016, Martinez2018, Foletto2020, Lunghi2015, Anwer2020, Tavakoli2020b, Farkas2021}).  Typically, these considerations take place in prepare-and-measure scenarios, i.e.~experiments in which a sender communicates quantum systems and a receiver measures them.

However, assigning a fixed finite dimension to a real-world quantum system is often an idealisation. Typically, it is an approximation of what is actually an infinite-dimensional system. Common platforms for qubit communication, such as weak coherent pulses or polarisation photons obtained by spontaneous parametric down-conversion constitute relevant examples. Indeed both very nearly correspond to harmonic oscillator qubits and polarisation qubits respectively, but the former still features higher-order oscillations and the latter still features multi-photon emissions. Whereas such dimensional deviations may often be viewed as neglegible noise in device-dependent protocols, it is much less clear whether the same is true in semi-device-independent quantum information protocols, namely when experimental devices are mostly uncharacterised. In fact, the practical challenges associated to a precise quantum dimension have in recent times partly motivated semi-device-independent concepts that are entirely different from the quantum dimension \cite{Chaves2015, VanHimbeeck2017, Wang2019, Zambrini2020, Zambrini2022, Tavakoli2021b}.

Here, we aim to remedy the shortcomings of dimension-based semi-device-independent quantum information protocols while maintaining the basic interest in the quantum dimension. To this end, we introduce and investigate systems that only nearly admit a faithful description in terms of qudits. These ``almost qudits'' are formulated operationally, i.e.~in a platform-independent way, and can thus be readily adapted to various quantum systems commonly modelled with a fixed dimension. We formalise the concept in the ubiquitous prepare-and-measure scenario and demonstrate its relevance by revisiting two established dimension-based quantum information protocols, for random number generation \cite{Li2011, Li2012} and  for  certification of multi-outcome measurements \cite{Tavakoli2020b} respectively, and showcase how tiny higher-dimensional contributions can in some cases significantly compromise their conclusions. Small deviations from the assumed quantum dimension can cause compromised security for random number generation and false positives for measurement certification. These observations motivate us to develop general tools for analysing almost qudit correlations. We introduce a hierarchy of semidefinite programming relaxations for bounding the set of almost qudit quantum correlations. We demonstrate its usefulness by fully resolving the issues observed for the two dimension-based  protocols. Then, we change perspective and consider almost qudits as a resource for the experimenter; we show how to control the higher-dimensional components in order to optimally boost the performance of the Quantum Random Access Code \cite{Ambainis2002}. Lastly, we discuss how our semidefinite programming hierarchy constitutes a general and useful tool for the well-researched task of  device-independently testing the dimension of a physical system.

\textit{Almost qudits in the prepare-and-measure scenario.---} A qu$d$it is a quantum state that can be represented by a  density matrix in a Hilbert space of dimension $d$, i.e.,~$\rho\in\mathcal{D}(\mathbb{C}^d)$. We say that quantum states in an experiment can be described by \emph{almost qudits} $\rho$, if the states are in principle require a representation in a countably unbounded Hilbert space ($\rho\in\mathcal{D}(\mathbb{C}^D)$ for any $D\geq d$), but their support is almost entirely on a $d$-dimensional subspace. Formally, we require that it is possible to choose a representation such that
\begin{equation}\label{almostqudit}
\Tr\left(\rho \Pi_d\right)\geq 1-\epsilon \, ,
\end{equation}
for all states $\rho$ where $\Pi_d=\sum_{j=1}^d \ketbra{j}{j}$ is the projector onto the qudit subspace and $\epsilon\in[0,1]$ is a deviation parameter quantifying the failure to admit a qudit description. 
The limiting cases, $\epsilon=0$ and $\epsilon=1$, correspond to a standard qudit and to an arbitrary quantum state respectively. Thus, we are mainly interested in the regime $0<\epsilon\ll 1$.  As a simple example, the optical coherent state $\ket{\alpha}=e^{-\frac{|\alpha|}{2}}\sum_{n=0}^\infty \frac{\alpha^n}{\sqrt{n!}}\ket{n}$ has $D=\infty$ in the Fock basis but for small average photon numbers (i.e.~$|\alpha|\ll 1$) it corresponds to an almost qubit ($d=2$) with $\epsilon=1-e^{-|\alpha|}(1+|\alpha|^2)\approx |\alpha|$.

We compare the almost qudit condition \eqref{almostqudit} to the trace norm condition $\norm{\rho - \Pi_d \rho \Pi_d}_1 \leq \epsilon'$. In general, the trace-norm condition is stronger than the almost qudit constraint. Indeed, the former implies the latter, as
\begin{equation}
1- \tr(\rho \Pi_d)  \leq \tr \abs{\rho-\Pi_d \rho \Pi_d} = \norm{\rho-\Pi_d \rho \Pi_d}_1 \leq \epsilon' \,.
\end{equation}
Unless the operator $\Pi_d \rho \Pi_d$ is positive, the converse is false. Winter's gentle measurement lemma \cite{Winter1999} implies simple upper bound on the trace distance between $\rho$ and $\Pi_d \rho \Pi_d$ given the almost qudit condition \eqref{almostqudit}, namely
\begin{equation}
\norm{\rho-\Pi_d \rho \Pi_d}_1 \leq \epsilon' \equiv \sqrt{8 \epsilon} \,. \label{eq:gentle}
\end{equation}

The trace-norm condition has the operational interpretation that there exists no experimental procedure by which an almost qudit can be distinguished from its (unnormalised) qudit projection with an accuracy greater than $\epsilon'$.

More generally, consider a prepare-and-measure experiment featuring a sender, Alice and a receiver, Bob. Alice selects an input $x\in\{1,\ldots,n_X\}$ and is assumed to prepare a qudit state $\rho_x$ that is sent to Bob, who in turn selects an input $y\in\{1,\ldots,n_Y\}$ and performs a corresponding quantum measurement $\{M_{b|y}\}_b$ with outcome $b$. The correlations are 
\begin{equation}\label{born}
p(b|x,y)=\Tr\left(\rho_x M_{b|y}\right) \,.
\end{equation}

If the states $\rho_x$ in the experiment are not exactly qudits, but only almost qudits \eqref{almostqudit} associated to the deviation parameters $\epsilon_x$, the probabilities can change by at most $\abs{p^{(\epsilon_x)}(b|x,y) - p^{(0)}(b|x,y)} \leq 2\epsilon'_x = 4\sqrt{2\epsilon_x}$. In Appendix~\ref{AppUniversalBound} we also show that for any linear functional $W=\sum_{bxy} c_{bxy} p^{(\epsilon_x)}(b|x,y)$, for real coefficients $c_{bxy}$, the maximal value based on Alice preparing almost qudits ($W^{(\epsilon_x)}$) can be bounded by a perturbation of the maximal value associated to standard qudits ($W^{(0)}$), namely
\begin{equation}\label{perturb}
W^{(\epsilon_x)}\leq W^{(0)} + 2 \sum_{xy}\epsilon'_x  \max_b \abs{c_{bxy}}.
\end{equation}
Since the correction is of order $\max_x \epsilon'_x$, one might believe that the practical impact of almost qudits on dimension-based quantum information protocols is accordingly small, and that such a perturbative approach would  for practical purposes suffice for their analysis. However, as we will see explicitly, such intuition is misguided. A more sophisticated analysis is needed to remedy the limitations of dimension-based protocols without rendering their success rates considerably sub-optimal or even vanishing. 

Finally, as with dimension-based correlations but unlike some other prepare-and-measure frameworks \cite{Wang2019, Tavakoli2021b}, almost qudit correlations have a natural classical analog. The classical case corresponds to assuming that all states are diagonal in the same basis, i.e.~$\rho_x=\sum_{m} p(m|x)\ketbra{m}{m}$. The assumption \eqref{almostqudit} simplifies to $\forall x: \sum_{m=1}^d p(m|x) \geq 1-\epsilon_x$. It follows that the set of classical correlations is a polytope. Without loss of generality, it can be characterised using a finite alphabet for $m$ by following the methods of \cite{Zambrini2022}.

\textit{Impact of almost qubits on random number generation.---} We investigate the magnitude of the impact of tiny higher-dimensional contributions on a well-known qubit-based protocol for random number generation \cite{Li2011, Li2012}. The protocol relies on the Quantum Random Access Code in the scenario $(n_X,n_Y,n_B)=(4,2,2)$, where Alice's input is represented as two bits $x_1$ and $x_2$: Bob randomly selects one of them which he aims to recover. On average, the probability of success reads $p_\text{RAC}=\frac{1}{8}\sum_{x_1,x_2=0,1}\sum_{y=1,2} p(b=x_y|x,y)$. When Alice sends qubits, the optimal quantum protocol achieves $p_\text{RAC}^\text{Q}=\frac{2+\sqrt{2}}{4}$. The protocol uses $p_\text{RAC}$ as a security parameter to certify that $b$ is random (e.g.~when $(x,y)=(1,1)$) also for an adversary who controls the devices via classical side information $\lambda$. The randomness can be quantified by the conditional min-entropy $R=-\log_2\left(P_g\right)$, where $P_g$ is the largest probability of guessing $b$, i.e.~$P_g=\max\{p(1|1,1),p(2|1,1)\}$, compatible with the observed value of $p_\text{RAC}$.

\begin{figure}[t!]
	\centering
	\includegraphics[width=0.95\columnwidth]{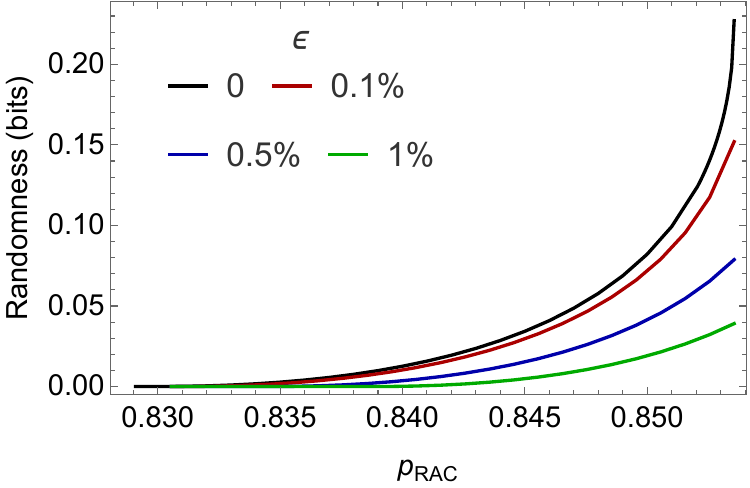}
	\caption{Randomness certified by the observed parameter $p_\text{RAC}$ for different deviation parameters $\epsilon$. The black curve corresponds to a standard qubit-based protocol.}\label{fig_randomness}
\end{figure}

Consider for simplicity a perfect value $p_\text{RAC}=p_\text{RAC}^\text{Q}$, which gives $R=-\log_2\left(p_\text{RAC}^\text{Q}\right)\approx 0.228$ bits of randomness \cite{Li2011} under a qubit assumption. However, the amount of certified randomness reduces considerably if the physical implementation uses almost qubits. For instance, choosing only $\epsilon_x=10^{-3}$, we numerically found via a seesaw procedure a much less random quantum model, implying the upper bound  $R\lesssim 0.152$ bits. Thus, a 1\textperthousand  \, deviation from a faithful qubit leads to a standard qubit-based analysis overestimating the randomness by at least about $50\%$.  Playing the role of the adversary, we systematically searched numerically for quantum models for some small choices of $\epsilon$ with the aim of maximally compromising the amount of certified randomness. The results are illustrated in Fig.~\ref{fig_randomness}. We see that the amount of certified randomness drops rapidly with $\epsilon$ and that the detrimental impact is largest for well-performing experiments that manage to approach the optimal value $p_\text{RAC}^\text{Q}$.

\textit{Impact of almost qubits on measurement certification.---} As a second example, we consider the impact of almost qubits on a qubit-based protocol for certifying genuine four-outcome measurements. In Ref.~\cite{Tavakoli2020b}, such a scheme is reported in the scenario $(n_X,n_Y)=(4,4)$ where the first three measurement settings have binary outcomes ($b\in\{1,2\}$) but the fourth setting has four possible outcomes ($b\in\{1,2,3,4\}$). The task corresponds to the following objective,
\begin{align}
\mathcal{A}\equiv \frac{1}{12}\sum_{x=1}^4\sum_{y=1}^3 p(t_{x,y}|x,y)-\frac{1}{5}\sum_{x=1}^4 p(x|x,y=4),
\end{align}
where $t=[1,1,1; \hspace{1mm}1,2,2; \hspace{1mm}2,1,2; \hspace{1mm}2,2,1]$. The optimal value for qubits is $\mathcal{A}^\text{Q}=\frac{3+\sqrt{3}}{6}\approx 0.7887$. To achieve this, the setting $y=4$ must correspond to a qubit SIC-POVM. It was proven that $\mathcal{A}\gtrsim 0.78367$ implies that $y=4$ corresponds to a genuine four-outcome measurement, i.e.~a measurement that cannot be reduced to a classical mixture of measurements with at most three outcomes. This was experimentally certified by observing $\mathcal{A}\approx 0.78514$ \cite{Tavakoli2020b}.

\begin{figure}[t!]
	\centering
	\includegraphics[width=0.95\columnwidth]{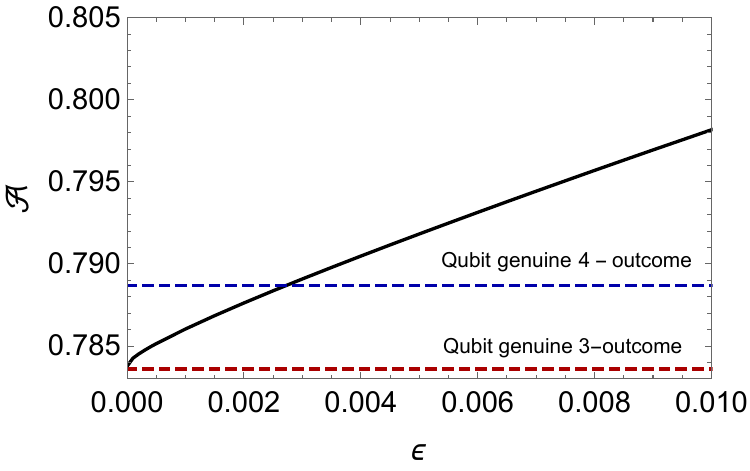}
	\caption{Correlation function $\mathcal{A}$ versus the deviation parameter $\epsilon$ for almost qubits with ternary-outcome measurements (full black line). Dashed lines are ternary (red) and quaternary (blue) bounds on $\mathcal{A}$ assuming perfect qubits.  }\label{fig_sic}
\end{figure}

Using a seesaw routine, we numerically found an almost qubit model with deviation parameter $\epsilon_x\approx 5 \times 10^{-4}$ that reproduces the experimentally observed certificate using only a ternary-outcome measurement. This would constitute a false positive when the lab states are not exactly faithful qubits. Moreover, using only $\epsilon_x\approx 3\times 10^{-3}$, ternary-outcome measurements can even exceed the qubit quantum limit $\mathcal{A}^\text{Q}$. These results are part of the systematic numerical search, see Fig~\ref{fig_sic},  for the trade-off between $\mathcal{A}$ and $\epsilon$ for ternary-outcome measurements. 

Finally, in Appendix~\ref{App:selftest}, we also investigate the impact of almost qubits on self-testing protocols based on the Quantum Random Access Code \cite{Tavakoli2018, Farkas2019, Mohan2019, Miklin2020}. It quantitatively benchmarks a preparation device that aims to emit the four states used in the BB84 quantum key distribution protocol.

\textit{Semidefinite relaxations.---} The considerable impact of almost qudits in dimension-based quantum information tasks naturally motivates the development of methods for analysing the set of almost qudit correlations. We introduce a hierarchy of semidefinite programming relaxations for bounding this set in arbitrary prepare-and-measure scenarios. This consists of a sequence of computable necessary conditions for the existence of an almost qudit model for a given distribution $p(b|x,y)$.

Define $S=\{\openone, V,\rho_1,\ldots,\rho_{n_X},M_{1|1},\ldots,M_{n_B|n_Y}\}$ where $\openone$ is the identity on $\mathbb{C}^D$ and $V$ is an auxiliary operator whose properties are to be specified. We can w.~l.~g assume that $\rho_x$ is pure ($\rho_x=\rho_x^2$), see Appendix \ref{App:purestates}. Also, we can w.~l.~g assume that the measurements are projective ($M_{b|y}M_{b'|y}=\delta_{b,b'}M_{b|y}$) because of the possibility of Neumark dilations. 
Build a monomial list $\mathcal{S}$ which consists of products of the elements of $S$. The choice of which products to include is a degree of freedom and corresponds to the level of the relaxation. Then, associate a $|\mathcal{S}|\times |\mathcal{S}|$ moment matrix 
\begin{equation}\label{moment}
	\Gamma_{u,v}=\Tr\left(uv^\dagger\right),
\end{equation}
for $u,v\in \mathcal{S}$. Importantly, the quantum probabilities \eqref{born} appear as elements in $\Gamma$ and are therefore fixed to the values $p(b|x,y)$. Due to rules such as normalisation of states, cyclicity of trace and projectivity of measurements, many elements in $\Gamma$ are equivalent. The remaining entries are viewed as free variables. By construction $\Gamma$ is postive semidefinite. 

Next, we impose the almost qudit property. To this end, we make use of the unphysical operator $V$ to emulate the projection operator $\Pi_d$. Thus, we insist that $V$ is projective ($V=V^2$) and that its trace is $d$ ($\Tr V=d$). The former impacts the equivalences among the entries of $\Gamma$ while the latter implies the additional constraint $\Gamma_{\openone,V}=d$. The almost qudit constraint \eqref{almostqudit} can then be imposed through explicit constraints on $\Gamma_{\rho_x,V}$. A necessary condition for the existence of a quantum model is the feasibility of the following semidefinite program, 
\begin{align}\nonumber
 \text{find } \Gamma \quad \text{ s.t. } \quad&  \forall x:\hspace{3mm} \Gamma_{\openone,\rho_x}=1, \quad \Gamma_{\openone,V}=d \\
&  \Gamma_{\rho_x,V}\geq 1-\epsilon_x, \quad \text{and} \quad \Gamma\succeq 0 .  \label{optQ}
\end{align}
Furthermore, this tool can be immediately adapted to bounding the maximal quantum value of a generic linear objective function: simply substitute the feasibility problem \eqref{optQ} for a maximisation problem in which $\Gamma_{\rho_x,M_{b|y}}$ are now free variables compounding the objective function.

\textit{Almost qudit protocols.---} We showcase the practical usefulness of the semidefinite relaxation hierarchy by applying it to the previously considered protocols. Already in Fig~\ref{fig_randomness}, we reported upper bounds on the randomness under the almost qubit assumption. In order to be able to certify randomness from almost qubits, we require lower bounds. It is well-known that upper bounds on the guessing probability $P_g$ (lower bounds on $R$) under quantum correlation constraints are typically compatible with semidefinite relaxations \cite{Masanes2011}. Using our method with a moment matrix of size $115$ we have reproduced the curves in Fig~\ref{fig_randomness} up to solver precision. Thus, these curves constitute the optimal randomness extraction for almost qubits. This can be compared to the method based on naive use of perturbations of a standard qubit scenario, following Eq.~\eqref{perturb}. In Appendix~\ref{AppPerturbative}, we perform the perturbative analysis for the randomness generation and find that already for $\epsilon>10^{-5}$, randomness cannot be certified at all.

Similarly, using a moment matrix of size 235 we are able to prove that the previously  reported value of $\epsilon$ for a falsely positive genuine four-outcome measurement in fact is optimal. More generally, we obtain tight upper bounds on the maximal value of $\mathcal{A}$ for any $\epsilon$  under ternary-outcome measurements. These accurately coincide with the lower bounds numerically reported in Fig~\ref{fig_sic}. Thus, the certification can now be performed under the almost qubit assumption. In comparison, performing the same analysis using the perturbative approach \eqref{perturb} leads to significantly suboptimal bounds (see Appendix~\ref{AppPerturbative}). For example, under quaternary-outcome measurements, a perturbative analysis deduces a deviation parameter $\epsilon\approx 3 \times 10^{-8}$ from the experimental value of $\mathcal{A}$ in \cite{Tavakoli2020b}. However, the optimal deviation parameter, pinpointed through semidefinite relaxations, is $10^{5}$ times larger.

\textit{Almost qubits as a resource.---} So far, we have considered situations in which the experimenter aims to prepare a qudit but fails to control the small higher-dimensional components of the lab state. Consider now the complementary situation in which the experimenter  has the ability to manipulate the entire almost qudit system. Then, almost qudits become a resource for boosting quantum communication beyond standard qudits. An interesting demonstration of this is obtained from Alice sending the following four states 
\begin{align}\nonumber
&\ket{\phi_{00}} = \sqrt{1-\epsilon}\ket{0} + \sqrt{\epsilon}\ket{2},  &&\ket{\phi_{10}} = \sqrt{1-\epsilon}s^+_{01} - \sqrt{\epsilon}s^+_{23}, \\\nonumber
&\ket{\phi_{11}} = \sqrt{1-\epsilon}\ket{1} + \sqrt{\epsilon}\ket{3}, &&\ket{\phi_{01}} = \sqrt{1-\epsilon}s^-_{01} - \sqrt{\epsilon}s^-_{23} ,
\end{align}
where $s^\pm_{ij}=\frac{\ket{i} \pm \ket{j}}{\sqrt{2}}$. These allow for boosting the success probability of the celebrated Quantum Random Access Code \cite{Ambainis2002}. By optimally choosing  the measurement operator $\{M_{0|y}\}$ as the projector onto the positive eigenspace of the operator $O_y=\sum_{x_1,x_2}(-1)^{x_y}\ketbra{\phi_{x_1x_2}}{\phi_{x_1x_2}}$, one finds the success probability 
\begin{equation}
p_\text{RAC}(\epsilon)=\frac{1}{2}+\frac{1}{4}\left[\sqrt{1+h(\epsilon)}+\sqrt{1-h(\epsilon)}\right] \, , \label{eq:pRACalmostqubit}
\end{equation} 
where $h(\epsilon)=(1-2\epsilon)\sqrt{1+4\epsilon-4\epsilon^2}$, which is valid for $\epsilon\leq \frac{1}{2}$.  A natural question is whether this is the best allowed by quantum theory. We have answered this in the positive by employing a semidefinite relaxation of size 107. For the most relevant case of small $\epsilon$, there is an immediate connection to the standard qubit scenario: the first-order approximation is $p_\text{RAC}(\epsilon)\approx \frac{2+\sqrt{2}}{4}+\frac{\epsilon}{\sqrt{2}}$, which is a linear correction to the success probability $p_\text{RAC}^\text{Q}$ of the standard Quantum Random Access Code. Note that a perturbative approach \eqref{perturb} would have overestimated the correction term by putting it at  $4\sqrt{2\epsilon}$.

\begin{figure}[t!]
	\centering
	\includegraphics[width=0.90\columnwidth]{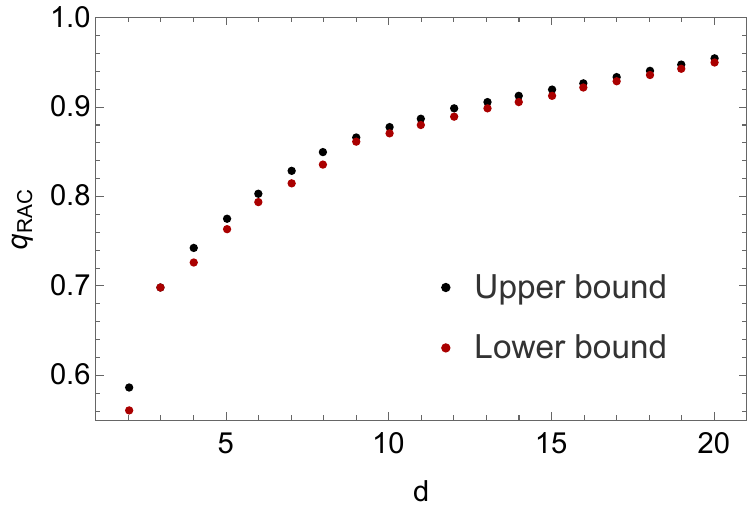}
	\caption{Upper and lower bounds on the success probability of the three-trit Quantum Random Access Code for qudits of dimensions $d=2,\ldots,20$. Upper bounds were computed using partially symmetrised semidefinite relaxations (variable elimination methods but not block-diagonalisation methods) at level 2 of the hierarchy. Lower bounds were computed by numerical search.}\label{fig_bigRAC}
\end{figure}

\textit{Bounding standard qudit correlations.---} An important special case of our semidefinite relaxation method is when $\epsilon_x=0$, corresponding to standard qudits. Naturally, bounding qudit correlations has been the subject of prior research \cite{Mironowicz2014, Navascues2015, Navascues2015b, Rosset2019}. The leading established method is also based on semidefinite relaxations \cite{Navascues2015} but differs significantly from ours. While \cite{Navascues2015} requires numerical sampling to construct the moment matrix, ours is fully deterministic. Also, although not strictly necessary, it typically favours separate semidefinite programs for all rank combinations of the measurement operators \cite{Navascues2015b}. This scales very quickly in all three parameters $(n_Y,n_B,d)$. In contrast, our method requires only a single semidefinite program. A key distinguishing feature of our method is that the complexity of the program is independent of $d$. Furthermore, it also applies to the classical case, relevant when linear programming becomes too expensive, simply by imposing  commutation constraints $[\rho_x,\rho_{x'}]=0$ and $[M_{b|y},M_{b'|y'}]=0$ in the moment matrix \eqref{moment}.  The main drawback is that our method does not converge to the quantum set of correlations (see Appendix~\ref{App:purestates} for an example). The basic reason is that our method equally well applies to a superset of qudit systems, namely correlations obtained from systems whose dimension, when averaged over a hidden variable, is $d$ \cite{Navascues2015b, Gribling2018}. Although convergence is also not known for the established method \cite{Navascues2015b}, there are cases in which it performs better.

We exemplify the usefulness of our method by addressing intermediate-scale dimensions in the simplest variant of a Quantum Random Access Code for which no analytical solution is presently known.  Alice has three trits $x_1x_2x_3\in\{1,2,3\}$ and communicates a $d$-dimensional system. Bob has one trit $y\in\{1,2,3\}$ and  aims to output $b=x_y$. The success probability is $q_\text{RAC}=\frac{1}{81}\sum_{x_1x_2x_3y}p(b=x_y|x,y)$. Invoking the standard symmetries of the Random Access Code (see \cite{Rosset2019}) to reduce the number of independent variables, we used semidefinite relaxations of size 1128 to bound  $q_\text{RAC}$ for every $d=2,\ldots,20$. Crucially, because the complexity of the computation is independent of $d$, we can readily evaluate also the higher-dimensional cases. 
In Fig.~\ref{fig_bigRAC} we plot the resulting upper bounds together with numerical lower bounds on $q_\text{RAC}$ obtained via an alternating convex search. These bounds are not expected to be optimal, but we conclude from the narrow difference between the upper and lower bounds that our semidefinite relaxations are at worst only nearly optimal. Importantly, we see that the gap tends to narrow even further as the dimension increases. This attests to the accuracy of the semidefinite relaxation method on the scale when it is most relevant, namely for higher dimensional systems.

\textit{Discussion.---} We have introduced almost qudits as an avenue to remedy the practical shortcomings of dimension-based quantum information protocols. We presented several examples showcasing the relevance of the concept; demonstrating how tiny deviations from an assumed dimension can significantly compromise the conclusions of established protocols. This led us to develop  methods for analysing almost qudits in the prepare-and-measure scenario and exemplify their usefulness both to concrete almost qudit problems as well as to established standard qudit problems.

Our work leaves several natural questions. Which experimental platforms are most and least prone to dimensional deviations from established theoretical models? What resources could an eavesdropper use to efficiently hack them? How do we wisely tailor protocols to perform well for almost qudit systems? These matters are particularly relevant in the context of the increasing interest in high-dimensional quantum information \cite{Bouchard2017, Erhard2018, Erhard2020, Chi2022, Ringbauer2022}. Moreover, what is the magnitude of quantum advantage possible from almost qudits, as compared to classical almost dits? Is it possible to find a converging hierarchy of semidefinite relaxations for characterising almost qudit correlations? Can these ideas be leveraged to semi-device-independent quantum key distribution protocols based on qubit assumptions \cite{Pawlowski2011, Woodhead2015}?

\underline{Note added:} 
A previous version of this manuscript incorrectly stated that the almost qudit constraint is equivalent to the trace-norm condition $\norm{\rho-\Pi_d \rho \Pi_d}_1 \leq \epsilon$  with the same $\epsilon$ as in Eq.~\eqref{almostqudit}, instead of the $\epsilon'$ of Eq.~\eqref{eq:gentle}.

\begin{acknowledgements}
 A. T.~is supported by the Wenner-Gren Foundations. J.P. and S.P. are supported by the Fonds de la Recherche Scientifique – FNRS under Grant No PDR T.0171.22 and under grant Pint-Multi R.8014.21 as part of the QuantERA ERA-NET EU program, by the FWO and the F.R.S.-FNRS under grant 40007526 of the Excellence of Science (EOS) program. J. P. is a FRIA grantee and S.P. is a Senior Research Associate of the Fonds de la Recherche Scientifique - FNRS. 
\end{acknowledgements}

\bibliography{references_almostQudits}

\onecolumngrid

\newpage

\appendix

\section{Perturbative bound for almost qudit correlations}\label{AppUniversalBound}
Given any linear function of the probabilities,
\begin{equation} 
	W = \sum_{bxy} c_{bxy} p(b|x,y) \,,
\end{equation}  we show that the quantum maximum for an almost qudit strategy is subject to the universal bound
\begin{equation}\label{unibound}
	W^{(\epsilon_x)} \leq W^{(0)} + 2 \sum_x \epsilon'_x \sum_{y} \max_b \abs{c_{bxy}} \, ,\end{equation}
	where $\epsilon' \equiv \sqrt{8 \epsilon}$ and $W^{(\epsilon_x)}$ is the maximum quantum value of $W$ for almost qudits with deviation parameters $ \epsilon_x $,
	\begin{equation}
	W^{(\epsilon_x)} \equiv \max_{\rho_x, M_{b|y}} W^{(\epsilon_x)} \qty(\{ \rho_x, M_{b|y} \}) \equiv W^{(\epsilon_x)} \qty(\{ \rho_x', M_{b|y}' \}) \, . 
\end{equation}  

 To see this, remember that any set of almost qudit states $\rho_x$ is $\epsilon' \equiv \sqrt{8 \epsilon}$-close to their (unnormalised) projection on the qudit space, $\norm{\rho_x- \Pi_d \rho_x \Pi_d }_1 \leq \epsilon' \equiv \sqrt{8 \epsilon}$. Defining their normalised projections $\tilde{\rho}_x = \frac{\Pi_d \rho_x \Pi_d }{\norm{\Pi_d \rho_x \Pi_d }}$, we use the triangle inequality to get
 
 \begin{equation}
 	\norm{\rho_x - \tilde{\rho}_x} \leq \ \norm{ \rho_x - \Pi_d \rho_x \Pi_d} + \norm{\Pi_d \rho_x \Pi_d - \tilde{\rho}_x} \leq 2\epsilon'_x \, ,
 \end{equation}
i.e., any almost qudit state is $2\epsilon'_x$-close to its normalised projection on the qudit space. 

For any set of measurements $\{M_{b|y} \}$, we find
\begin{align}
	W^{(\epsilon_x)}(\{\rho_x, M_{b|y} \}) - W^{(0)}(\{\tilde{\rho}_x,M_{b|y} \}) &= \sum_{bxy} c_{bxy}  \Tr \qty[ \qty(\rho_x - \tilde{\rho}_x ) M_{b|y} ] \\
	&= \sum_x \Tr \qty[ \qty( \rho_x - \tilde{\rho}_x) \sum_{by} c_{bxy} M_{b|y} ]\\
	&\leq \sum_{xy}  \norm{ \rho_x - \tilde{\rho}_x} \norm{\sum_b c_{bxy} M_{b|y} }_\infty \\
	& \leq 2 \sum_x \epsilon'_x \sum_{y} \norm{\sum_{b} c_{bxy} M_{b|y}}_\infty \\
	& \leq 2 \sum_x\epsilon'_x \sum_{y} \max_b \abs{c_{bxy}} \, ,
\end{align}
where in the fourth line we used H\"older's inequality and in the last line we use that $\sum_b M_{b|y}= \openone$.

The above inequality applies in particular to the almost qudit strategy ($\{\rho_x', M_{b|y}'\}$) that realises the maximal value  $W^{(\epsilon_x)}$. Thus, we have the following chain of inequalities, 
\begin{equation}
	W^{(\epsilon_x)} \leq W^{(0)} \qty(\{\tilde{\rho}_x',M_{b|y}' \}) +  2  \sum_{xy} \epsilon'_x\max_b \abs{c_{bxy}} \leq  W^{(0)} + 2 \sum_{xy}\epsilon'_x  \max_b \abs{c_{bxy}} \, .\end{equation}

\section{Self-testing and almost qubits} \label{App:selftest}

The Quantum Random Access Code is also a benchmark for self-testing protocols. Under a qubit restriction, an observation of the optimal value $p_\text{RAC}^\text{Q}$ certifies that the preparation device emits the four states used in the BB84 QKD protocol. That is, up to a global unitary, the preparations are given by $\ket{\psi_{x_1x_2}}=H^{x_1}Z^{x_2}\ket{0}$ where $H$ is the Hadamard gate. The quality of the preparation device can then be quantified based on the smallest average fidelity between the BB84 states and the possible lab states compatible with the observed value of $p_\text{RAC}$, i.e.~
\begin{equation}
\mathcal{F}=\min_{\{\rho_{x_1x_2}\}\in T(p_\text{RAC})} F_\text{avg},
\end{equation}
where the average fidelity 
\begin{equation}
F_\text{avg}=\max_\Lambda \frac{1}{4}\sum_{x_1,x_2=0,1} \bracket{\psi_{x_1x_2}}{\Lambda[\rho_{x_1x_2}]}{\psi_{x_1x_2}}
\end{equation}
itself is optimised over an arbitrary  extraction channel $\Lambda$.

\begin{figure}
	\centering
	\includegraphics[width=0.5\columnwidth]{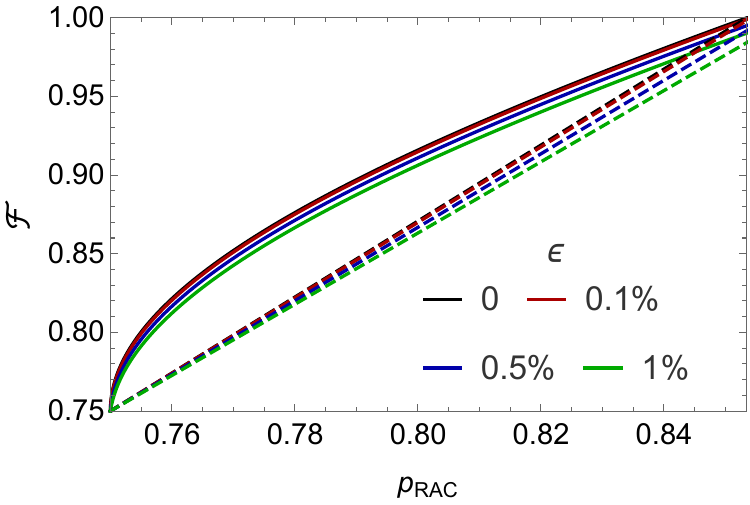}
	\caption{The average fidelity $\mathcal{F}$ of the prepared states with the BB84 ensemble $\ket{\psi_{x_1x_2}}=H^{x_1}Z^{x_2}\ket{0}$ in function of the observed value $p_\text{RAC}$ for different deviation parameters $\epsilon$. Results with and without shared randomness are plotted in dashed and full lines respectively. }\label{fig_self_test}
\end{figure}

Consider now the impact of an $\epsilon$-deviation from the qubit assumption. For instance, let the device prepare almost qubits
\begin{align}\nonumber
&\ket{\phi_{00}} = \sqrt{1-\epsilon}\ket{0} + \sqrt{\epsilon}\ket{2},  &&\ket{\phi_{10}} = \sqrt{1-\epsilon}s^+_{01} - \sqrt{\epsilon}s^+_{23}, \\\label{racstates}
&\ket{\phi_{11}} = \sqrt{1-\epsilon}\ket{1} + \sqrt{\epsilon}\ket{3}, &&\ket{\phi_{01}} = \sqrt{1-\epsilon}s^-_{01} - \sqrt{\epsilon}s^-_{23} ,
\end{align}
where $s^\pm_{ij}=\cos \theta \ket{i} \pm \sin \theta \ket{j}$. The optimal success probability can be computed analytically as $p_\text{RAC}=\frac{1}{2}+\frac{1}{8}\sum_y \lambda_+(O_y)$, by optimally choosing the measurement operator $\{M_{0|y}\}$ as the projector onto the positive eigenspace of the operator $O_y=\sum_{x_1,x_2}(-1)^{x_y}\ketbra{\phi_{x_1x_2}}{\phi_{x_1x_2}}$. Using the Choi representation of $\Lambda$, the average fidelity associated to the states can be computed as a semidefinite program for a given $\theta$ and $\epsilon$, which then immediately yields an upper bound on $\mathcal{F}$. In Fig.~\ref{fig_self_test} we display the results, both with and without shared randomness. Alike the case of random number generation, we see that a well-performing experiment, corresponding to larger values of $p_\text{RAC}$, is more sensitive to the impact of small higher-dimensional components. Numerically, we have found no strategy with a larger impact on the self-testing conclusions of the qubit-based protocol.

\section{Restriction to pure states \& non-convergence} \label{App:purestates}

Here, we show that we can restrict to pure states $\tilde{\rho}_x = \tilde{\rho}_x^2$ at the level of the hierarchy. That it, we show that for every strategy that generates the behaviour $p(b|x,y) = \tr\qty(\rho_x M_{b|y})$  with mixed states $\tilde{\rho}_x \ne  \tilde{\rho}_x^2$ measurements $M_{b|y}$, we can find an equivalent pure strategy $\tilde{\rho}_x = \tilde{\rho}_x^2$ with measurements $\tilde{M}_{b|y}$ that generates the same behaviour while still satisfying all SDP constraints in the relaxation.

One can always decompose a mixed state $\rho_x$ as a sum of pure states $\ket{\psi_\lambda^x}$\footnote{In general, the weights in the decomposition will depend on the state, i.e. in general, $\rho_x = \sum_\lambda p^x(\lambda) \ketbra{\psi_\lambda^x}$. However, this can always be rewritten as $\rho_x = \sum_{\lambda_1,\dots,\lambda_{n_{X}}} p(\lambda_1,\dots,\lambda_{n_X}) \ketbra{\psi_{\lambda_x}^x}$, where $p(\lambda_1,\dots,\lambda_{n_X}) = \prod_{x=1}^{n_X} p^x(\lambda)$.}
\begin{equation}
\rho_x = \sum_\lambda p(\lambda) \ketbra{\psi_\lambda^x} \, .
\end{equation}
We may define a new set of states $\tilde{\rho}_x$ and measurements operators $\tilde{M}_{b|y}$ that generate the same behaviour $p$, as follows:
\begin{align}
\tilde{\rho}_x &= \bigoplus_\lambda \ketbra{\psi_\lambda^x} \, , 
&\tilde{M}_{b|y} = \bigoplus_\lambda M_{b|y} \, .
\end{align}
We also introduce a new identity operator
\begin{equation}
\tilde{V} = \bigoplus_\lambda V \, .
\end{equation}
By redefining the (unnormalised) trace as
\begin{equation}
\tilde{\tr} (X) = \sum_\lambda p(\lambda) \tr(X_\lambda) \, ,
\end{equation}
where $X = \bigoplus_\lambda X_\lambda$ and $\tr$ is the regular (normalised) trace, one may readily check that our new set of operators indeed generate the same behaviour i.e. $p(b|x,y) = \tr \qty( \tilde{\rho}_x \tilde{M}_{b|y} )$, satisfy all original SDP constraints;  $\tilde{\tr}(\tilde{\rho}_x) = 1$ for all $x$, $\tilde{V}=\tilde{V}^2$ with $\tilde{\tr} (\tilde{V}) = d$ and $\tilde{\tr} \qty(\tilde{\rho}_x \tilde{V}) > 1-\epsilon_x$ for all $x$, while in addition all states are pure, $\tilde{\rho}_x^2 = \tilde{\rho}_x$.

Precisely because shared randomness is built into the hierarchy, it cannot converge in general, i.e.~it can only guarantee the dimension on average -- not in every round. For example, if the Quantum Random Access Code  is implemented for $d=3$, the true quantum limit is $p_\text{RAC}=\frac{5+\sqrt{5}}{8}$. Our relaxation cannot go below the upper bound $p_\text{RAC}\leq \frac{6+\sqrt{2}}{8}$. The reason is that this value can be obtained from uniformly mixing the optimal qubit strategy with the trivial four-dimensional strategy, namely $\frac{1}{2}\times p_\text{RAC}^\text{Q}+\frac{1}{2}\times 1=\frac{6+\sqrt{2}}{8}$. On average, the dimension is $d=3$. The fact that our method cannot go beyond this limit follows from an argument analogous to the above: one can construct a new trace based on separately applying the standard trace to a qubit and ququart block of a moment matrix. In this case, one defines a new set of pure states from the optimal qubit and quart strategies, i.e., $\tilde{\rho}_x = \rho_x^{\rm BB84} \oplus \ketbra{x}{x}$ and similarly for the measurements. One then constructs a new identity operator $\tilde{V} = \openone_2 \oplus \openone_4$, where $\openone_d$ is the usual identity in dimension $d$ and redefines the trace as $\tilde{\tr}(X) = \frac{1}{2} \tr(X_1) + \frac{1}{2} \tr(X_2)$.

\section{Application perturbative analysis to protocols}\label{AppPerturbative}

We apply the perturbative analysis to the problem of randomness certification based on the RAC discussed in the main text. Say one observes a given value of the security parameter $p_{\rm RAC}^{(\epsilon_x)}$ in an experiment where the states deviate from perfect qubits by $\epsilon_x$. How much randomness can still be certified? Based on the bound \eqref{unibound}, we know that in the qubit subspace, we must have at least $p_{\rm RAC}^{(0)}  \geq  p_{\rm RAC}^{(\epsilon_x)} - \frac{1}{2}\sum_x \epsilon'_x$. For a given setting $x^\ast$, this value certifies a certain minimal guessing  probability $P_g$ in the qubit subspace (and hence randomness $R^{(0)}=-\log_2(P_g^{(0)})$). To find the certifiable randomness based on almost qubits, however, we must bound the guessing probability for our almost qubit ensemble, leading to a second correction, $P_g^{(0)} \leq P_g^{(\epsilon_x)} + 2\epsilon'_{x^\ast}$.

Taken together, we find 
\begin{equation} \label{eq:universal_randomness}
R^{(\epsilon_x)}(p_{\rm RAC}^{(\epsilon_x)}) \geq \log \qty(2^{R^{(0)}(p_{\rm RAC}^{(\epsilon_x)} - \frac{1}{2}\sum_x \epsilon'_x)} + 2\epsilon'_{x^\ast} ) \, ,
\end{equation}	
where $R^{(0)}(p_{\rm RAC})$ is the randomness for perfect qubits given one observed a security parameter $p_{\rm RAC}$. The perturbative analysis suggests that we require $\epsilon < 10^{-5}$ in order to certify any randomness. 
In the same vein, the equation \eqref{unibound} can be immediately applied to the problem of measurement certification discussed in the main text. In this case, the suboptimality of the perturbative approach is even more drastic. Unless $\epsilon <10^{-8}$, no genuine qubit POVM can be certified ($10^5$ times smaller than the actual value). Both of these examples illustrate that the hierarchy method is essential to analysing almost qudit corrections.

\end{document}